\documentclass[12pt,preprint]{aastex}

\def\kms{\ifmmode{\rm km\thinspace s^{-1}}\else km\thinspace s$^{-1}$\fi}

\shorttitle{A Low-mass Spectroscopic Binary}
\shortauthors{Blake et al.}

\begin{document}


\title{A Spectroscopic Binary at the M/L Transition}
\author{Cullen H. Blake\altaffilmark{1,5}}
\altaffiltext{1}{Harvard-Smithsonian Center for Astrophysics, 60 Garden Street,  Cambridge, MA 02138; cblake@cfa.harvard.edu}

\author{David Charbonneau\altaffilmark{1,6}}

\author{Russel  J. White\altaffilmark{2}}
\altaffiltext{2}{University of Alabama in Huntsville, Physics Department, 301 Sparkman Drive, 201B Optics Building, Huntsville, AL 35899}

\author{Guillermo Torres\altaffilmark{1}}

\author{Mark S. Marley\altaffilmark{3}}
\altaffiltext{3}{NASA Ames Research Center, MS 245-3, Moffett Field, CA 94035 }

\author{Didier Saumon\altaffilmark{4}}
\altaffiltext{4}{Los Alamos National Laboratory, PO Box 1663, MS F663, Los Alamos, NM 87545 }

\altaffiltext{5}{Harvard Origins of Life Initiative Fellow}
\altaffiltext{6}{Alfred P. Sloan Research Fellow}

\begin{abstract}
We report the discovery of a single-lined spectroscopic binary with an Ultra Cool  Dwarf (UCD) primary with a spectral type between M8 and  L0.5. This system was discovered during the course of an ongoing survey to monitor L dwarfs for radial velocity variations and is the first known small separation ($a<1$ AU) spectroscopic binary among dwarfs at the M/L transition. Based on radial-velocity measurements with a typical precision of 300 m s$^{-1}$, we estimate the orbital parameters of this system to  be $P=246.73\pm0.49$ d, $a_{1}\sin{i}=0.159\pm0.003$ AU, $M_{2}\sin{i}=0.2062 (M_{1}+M_{2})^{2/3}\pm0.0034$ $M_{\sun}$. Assuming a primary mass of $M_{1}=0.08M_{\sun}$ (based on spectral type), we estimate the secondary minimum mass to be $M_{2}\sin{i}=0.054 M_{\sun}$. With future photometric, spectroscopic, and interferometric observations it may be possible to determine the dynamical masses of both components directly, making this system one of the best characterized UCD binaries known.

\end{abstract}

\section{Introduction}

Near infrared (NIR) surveys, such as 2MASS, DENIS, and SDSS (with its $z$-band capability), have resulted in a rapid increase in our knowledge of 
the properties of stars. This is particularly true for the 
 late M, L and 
T spectral types, collectively known as Utracool Dwarfs (UCDs, \citealt{kirkpatrick2005}). Today we know of more than 600  
 L and T dwarfs\footnote{http://www.dwarfarchives.org}. Despite these discoveries, we still do not have a clear  
understanding of how molecular cloud
material assembles itself into such relatively low mass objects and fundamental properties such as temperature, age and mass remain coarsely determined. Binary star systems are a crucial
tool for addressing both of these issues.
UCD binaries allow us to directly measure the masses, and possibly the radii, of these objects and constrain theoretical models of the structure and emergent flux (i.e. \citealt{stassun2006}).  Even without determining absolute ages, assuming that the components of a binary have identical ages can provide constraints on atmospheric models through estimates of the luminosity ratios. Different formation scenarios make varied predictions for the 
statistical properties of UCD binary systems (see \citealt{burgasser2007a} for an overview). As a result, studying the properties of UCD binaries may shed light on the formation mechanism of the entire class of objects. 

Although models of field L and T dwarfs have reached a high degree of
sophistication, model spectra, particularly for L dwarfs, are highly
dependent upon the assumed cloud model.  \citet{cushing2008}, for
example, found that changes in the assumed degree of cloud
sedimentation could alter the derived effective temperature, $T_{\rm eff}$, by up to several
hundred Kelvin and the log of the surface gravity, $\log g$, by 0.5 dex. A spectroscopic binary offers the prospect of constrained masses and coevality and would
provide excellent simultaneous constraints on the cloud model, object
masses, and effective temperatures \citep{marley2008}.

The binarity of UCDs has been studied both spectroscopically and with direct imaging. Relative to main sequence stars (i.e. \citealt{duquennoy1991}), the UCD binaries are more rare, lie at closer separations, and are more likely to have components with equal masses (\citealt {burgasser2007a} and references therein). Owing to the difficulties in detecting close, high contrast ratio
binaries, imaging surveys do not provide a clear picture of UCD binarity at separations less than 1 AU (see \citealt{reid2008}). Despite these difficulties, \citet{bouy2004}, \citet{lane2001}, and \citet{golimowski2007} have used high-resolution imaging to determine the orbit of young L or M dwarf binaries. The binaries with small separations are more readily detectable by searching for the radial
velocity signal due to the reflex motion of the primary star. Spectroscopic searches for UCD binaries have identified several binaries and candidate binaries \citep{joergens2007, kenyon2005,guenther2003, basri2006}, though the total number of observations of each object tends to be small. Recent work by \citet{burgasser2007b} has demonstrated that low-resolution spectra alone can be used to detect the faint companions  by searching for the subtle signature of the cooler object at the wavelengths of certain molecular features. 

 We presented results from a pilot study targeting nine field L dwarfs with the Phoenix instrument on Gemini-S in \citet{blake2007}. Here, we report the 
detection of a single-lined UCD spectroscopic binary (SB1) with an orbital separation of approximately 0.4 AU. 2MASSJ03202839$-$0446358 (hereafter 2M0320$-$04) was included in the catalog of nearby, cool stars presented by \citet{cruz2003} who spectroscopically classified it as a possible  M8 dwarf based on optical spectra. \citet{wilson2003} classified the object as an L0.5 dwarf based on low resolution NIR spectroscopy. In \S 2 we describe the high-resolution NIR spectroscopy of 2M0320$-$04,
in \S 3 we describe the modeling process used to extract radial velocities from 
these data, the fit of a Keplerian orbit to the radial velocities, and the search for the spectral lines of the secondary, and in \S 4 we describe the implications 
of this discovery for future studies of UCDs.

\section{Observations}

We observed 2M0320$-$04 with the NIRSPEC spectrograph \citep{mclean1998} on the Keck telescope on 14 nights between September 2003
and January 2007 as part of a program to monitor a large sample of L dwarfs for radial velocity variations. In all cases, we used the same instrumental setup. This setup was selected to cover the CO bandhead and R-branch features around
2.3$\micron$. This spectral region is rich in telluric absorption features due to methane. As described in the next section, these telluric features will serve as the wavelength reference for our radial velocity measurements. We used a 3 pixel (0.432$\arcsec$) slit, with the N7 blocking filter, to 
produce an approximate scale of 0.3 \AA~pixel $^{-1}$ and a resolution of $R=\lambda/{\Delta \lambda}\approx25000$. The extracted spectra contain 1000 pixels and cover 
the approximate spectral range 2.285$\micron$ to 2.318$\micron$. Exposures ranged from 900s to 1200s per nod position with adjustments made according to observing conditions in order to maintain approximately constant S/N. The data were gathered in nod pairs so as to facilitate the subtraction of sky emission lines. In total, we collected 16 nod pairs.

\section{Analysis}
After subtraction of the nod pairs to remove sky emission lines, we extracted the spectra  following
the optimal extraction procedures outlined in \citet{horne1986}. We modeled the extracted spectra following
a procedure similar to that described in \citet{blake2007}. Our model begins with two high-resolution template spectra: one for the Earth's atmosphere and one  for the L dwarf. The high-resolution ($5\times10^{-6} \micron$ pixel$^{-1}$) spectrum of the Earth's atmosphere is provided by \citet{livingston1991}. The high-resolution synthetic spectra of L dwarfs were computed as described in \citet{marley2002}, with a number of improvements to be described in a future publication.   The models apply the condensation cloud model of \citet{ackerman2001} with a  sedimentation parameter of $f_{\rm sed}=3$, corresponding to a moderate amount of  condensate settling.  The models used here have solar metallicity \citep{lodders2003}, use the opacities described in \citet{freedman2008}, and a fixed gravity
of $\log g=5$ (cgs) and cover a range of $T_{\rm eff}$ from 1200 to  2400$\,$K.  The synthetic spectra provide monochromatic fluxes spaced $4.2 \times  10^{-6}\,\mu$m apart. We convolve and re-sample the product of the telluric and L dwarf high-resolution spectra to generate the model that we then fit to the extracted 1-D spectra. Our model has several free parameters. The parameters related to the L dwarf are the projected rotation velocity of the L dwarf ($V\sin{I}$) where $I$ indicates the inclination of the rotation axis to the line of sight, its  $T_{\rm{eff}}$, and its radial velocity. The parameters related to the spectrograph are the PSF width, flux normalization, and  the wavelength solution (i.e. the mapping from wavelength to pixel position).

For the parameters of the L dwarf we first determined a best initial fit with $V\sin{I}=16.5\pm0.5$ km s$^{-1}$ using the  L dwarf spectral 
template with $T_{\rm{eff}}=2200$K using least squares fitting. This spectral template provided the best overall fit to the data and is also consistent with the L0 spectral type \citep{basri2000, golimowski2004}. We modeled the wavelength solution as a third-order polynomial, the overall flux normalization as a third-order polynomial, and the spectrograph PSF as a single Gaussian. The PSF is expected to be asymmetric at some level, and to vary across the spectrograph order. Spectral models that accommodate these subtleties may yield even higher precision than that achieved here. With the $V\sin{I}$ and  $T_{\rm{eff}}$ fixed the model has a total of ten free parameters. For each spectrum, we determined the best fit values for each parameter, including the radial velocity, using the \textit{AMOEBA} algorithm. An example of a NIRSPEC spectrum and best fit model is shown in Figure 1. The algorithm failed to converge on a solution for one out of 28 spectra, resulting in a total of 27 radial-velocity measurements. Visual inspection of this spectrum indicated a strong flux gradient across the order, likely due to a poor extraction. We used a bootstrap simulation to estimate the error on the individual radial velocities, similar to the technique described in \citet{blake2007}. 
We simulated NIRSPEC observations of L dwarfs with different S/N and radial velocities and used our modeling procedure to estimate our ability to recover the known radial velocity. Since the NIRSPEC data are of relatively high S/N ($\approx 60$ per pixel), and the V$\sin{I}$ is a modest 16.5 km s$^{-1}$, the results of our simulation indicate that the statistical errors on the radial velocities should be rather small ($\approx 60$ m s$^{-1}$). The measured radial velocities are listed in Table 1.

An initial search for periodicity in the radial velocities with the Lomb-Scargle Periodogram revealed a clear signal with a period $P\approx 250$d, prompting a more detailed 
analysis. The radial velocities covering nearly 5 cycles of the binary were fit
with a Keplerian orbital model using standard non-linear least-squares
techniques. The six parameters of this model are the period, $P$, the systemic velocity, $\gamma$, the radial velocity semi-amplitude, $K_{1}$, the eccentricity, $e$, the time of periastron passage, $T_{0}$, and the longitude of periastron, $\omega$. The scatter in the fit was found to be
significantly larger than the internal uncertainties, possibly as a
result of systematic errors. For the final fit we adjusted the
internal errors by adding 0.32~\kms\ in quadrature to the error estimates based on the bootstrap simulations, so as to force a
reduced $\chi^2$ value near unity. The resulting orbital elements are
listed in Table~2, wherein we also
 state our estimate of the projected semi-major axis of the orbit of
 the primary, $a_1\sin{i}$, and the observations and orbital fit are shown
 in Figure 2. The eccentricity ($e=0.065\pm0.016$) is small but significant at
the 4$\sigma$ confidence level. The radial velocity semi-amplitude of the primary, $K_{1}=7.02\pm0.12$ km s$^{-1}$, is 21 times the per-point measurement precision.

\section{Discussion and Conclusions}

Based on the estimated spectral type (M8 to L0.5) we can estimate the effective temperature of the primary of 2M0320$-$04 if we assume that the light from the secondary is negligible. \citet{golimowski2004} present estimates of $T_{\rm{eff}}$ as a function of spectral type and from this work we estimate $2200<T_{\rm{eff}}<2400$K  for an assumed age of 3 Gyr, in general agreement with our fits to the L dwarf templates. Using the models of \citet{baraffe2003} to relate $T_{\rm{eff}}$ to mass at an age of 3 Gyr, we estimate that the primary has a mass $0.075<M_{1}<0.081$M$_{\sun}$. We can use the mass function derived from the radial-velocity fitting procedure to estimate the mass of the secondary as a function of $\sin{i}$ where $i$ is the inclination of the system to our line of sight. If we assume $M_{1}=0.08$M$_{\sun}$, and that $\sin{i}\approx1$, then the mass of the secondary is $M_{2}=0.054M_{\sun}$. Using the models of \citet{baraffe2003} and the work of \citet{golimowski2004}, the secondary would have $T_{\rm{eff}}\approx 1350$K with a spectral type of approximately L7 to T3 at the assumed age of 3 Gyr. As $\sin{i}$ decreases, the total mass of the system increases, resulting in an observed spectrum that is not dominated by the L0 primary. This would be inconsistent with observations without requiring that the more massive component of this system be significantly under-luminous. In particular, values of $\sin{i}<0.7$ would result in a secondary star that is more massive, though less luminous, than the primary. While it is possible that $\sin{i} \sim 1$ and the system presents eclipses, the relatively small value of $(R_{1}+R_{2})/a\sim10^{-3}$ makes eclipses unlikely. Given the constraint $\sin{i}>0.7$ the actual probability of observable eclipses is likely somewhat higher.

While we have no direct measure of the age of the system, it is possible to constrain the age from its kinematics. Using the
distance estimate of 26.2$\pm$ 4.3 pc from \citet{cruz2003}, the proper motion of $0.678\pm0.038\arcsec$ year$^{-1}$ \citep{deacon2005}, and our
measurement of the radial velocity, we can calculate the space velocity and U,V,W velocity components following \citet{johnson1987}. Since the parallax of this system is not known, the large error on the distance estimate results in large errors on the kinematic estimates.
We find velocity components [U,V,W]=[$-$62.0$\pm9.3$,$-$38.6$\pm8.9$, $-$34.3$\pm7.2$] km s$^{-1}$ relative to the local standard of rest. Using the age-velocity relation from Eqn. 8 of \citet{wielen1977}  we estimate an age based on the W velocity of $\tau > 3.6$ Gyr (95$\%$ confidence).

It is interesting to consider the detectability of the spectral lines of the secondary in our spectra, which would allow us to estimate directly the mass ratio. Based on the models of \citet{baraffe2003}, the expected $K$-band flux ratio for objects of these masses is $\approx10$ at $\tau=1$Gyr and $\approx100$ at $\tau=5$Gyr. While detecting the secondary lines at the later age would be challenging, methods like TODCOR \citep{zucker1994} have been used to recover secondaries in systems with flux ratios $\approx 50$ (D. Latham; private communication). Following a method similar to TODCOR we searched for the spectral lines of the secondary as follows. Using the orbital solution presented in Table 2, we searched a grid in two parameters; the mass of the secondary and the $K$-band flux ratio. At each grid point, a second theoretical template spectrum with $T_{\rm{eff}}=1400$K was added to the fitting procedure described in Section 3. The V $\sin{I}$ of the secondary was assumed to be the same as that of the primary. During this process the radial velocity of the primary of 2M0320$-$04 was fixed to the value from the orbital solution, the radial velocity of the secondary was also fixed based on the orbital solution and the assumed value of $M_{2}$, and the $K$-band flux ratio was fixed. At each grid point this modified modeling scheme was used to determine the best fit of this model to the data. Significant improvements in $\chi^2$ from the case of a flux ratio of 0.0 would indicate the detection of the secondary component. We carried out this procedure for a subset of 12 of our spectra gathered near times of quadrature, when the velocity separation of the primary and secondary would likely be greatest. We found no evidence for the spectral signature of a secondary component with a flux ratio greater than 0.1 in $K$ band. We note that the non-detection of the secondary spectral lines also implies $\sin{i}\approx1.0$. If $\sin{i}=0.8$ the expected flux ratio at $\tau=5$ Gyr would be $\sim0.3$ and the secondary would likely have been detected. Detection of the spectral lines of the secondary and resolution of the system with direct imaging would allow for the first direct measurement of  the mass of a field T dwarf. Since the orbital solution combined with direct imaging provides a distance measurement,  future observations of this system could also provide important empirical tests of theoretical models for old objects at such low masses.

 This work demonstrates the importance of radial velocity searches for binary UCDs with small ($a<1$ AU) separations as a complement to the direct imaging searches for companions at larger separations. A more detailed analysis of our data, including a more sophisticated model of the individual spectra, may lead to the detection of the spectral lines of the secondary and an estimate of the mass ratio. If the inclination can also be measured, then the masses of both components may be determined. If we assume the models of \citet{baraffe2003}, $\sin{i}=1.0$, and a distance of 26 pc \citep{cruz2003}, then the maximum angular separation of the pair is $\approx 16$ mas, below the capabilities of the Keck Laser Guide Star AO system. While the flux ratio may be large ($\Delta K\approx 5$ mag, depending on age), future interferometric systems may be able to resolve both components and provide the measurement of the inclination required to directly measure the masses of both components.

\textit{Note:} During the completion of this Letter we became aware of work by \citet{burgasser2008} describing a tentative detection of the secondary
in the 2M0320$-$04 system using the spectral diagnostics described in \citet{burgasser2007b}.

\acknowledgments
CB acknowledges support from the Harvard Origins of Life Initiative. GT acknowledges partial support from NSF grant AST-0708229. We thank the referee for thoughtful comments that helped to improve this manuscript. Part of this research was supported by a {\it Spitzer Science
Center} Theory grant. The data presented herein were obtained at the W.M. Keck Observatory,
which is operated as a scientific partnership among Caltech, the
University of California, and NASA. The Observatory was made possible by
the generous financial support of the W.M. Keck Foundation. The authors wish to recognize and acknowledge the very significant cultural role and reverence that the summit of Mauna Kea has always had within the indigenous Hawaiian community.  We are most fortunate to have the opportunity to conduct observations from this mountain. This research has benefited from the M, L, and T dwarf compendium housed at DwarfArchives.org and maintained by C. Gelino, D. Kirkpatrick, and A. Burgasser.

{\it Facilities:} \facility{Keck/NIRSPEC}

\clearpage

\begin{deluxetable}{ccc}
\tabletypesize{\scriptsize}
\tablecaption{RV Data for 2M0320$-$04}
\tablewidth{0pt}
\tablehead{
\colhead{HJD-2400000} & \colhead{RV} & \colhead{$\sigma_{RV}$}\\
\colhead{} & \colhead{km s$^{-1}$} & \colhead{km s$^{-1}$}}
\startdata

  52921.0960  &    6.95  &     0.33  \\
    52921.1102  &    6.26  &   0.33  \\
    52922.1054  &    6.70  &   0.33  \\
    52922.1196  &    6.35  &   0.33  \\
    52957.0218  &    4.18  &   0.34  \\
    52957.0361  &    4.49  &   0.33  \\
    53272.1197  &   -6.17  &   0.33  \\
    53272.1304  &   -6.85  &   0.33  \\
    53273.0793  &   -6.56  &   0.33  \\
    53273.0900  &   -5.97  &   0.33  \\
    53328.8225  &   -3.97  &   0.34  \\
    53328.8343  &   -4.03  &   0.33  \\
    53421.7131  &    6.74  &   0.33  \\
    53421.7239  &    6.18  &   0.32  \\
    53669.8797  &    6.53  &   0.32  \\
    53669.8919  &    5.84  &   0.33  \\
    53670.8719  &    6.37  &   0.33  \\
    53670.8841  &    6.26  &   0.33  \\
    53686.8538  &    5.59  &   0.33  \\
    53686.8659  &    5.78  &   0.33  \\
    53742.8030  &   -3.25  &   0.33  \\
    53742.8151  &   -2.92  &   0.33  \\
    54023.9636  &   -7.25  &   0.33  \\
    54023.9757  &   -7.70  &   0.33  \\
    54100.7404  &    2.03  &   0.33  \\
    54100.7526  &    1.03  &   0.33  \\
    54101.7468  &    1.67  &   0.33  \\

\enddata
\tablecomments{Individual radial-velocity measurements. Errors include the estimated statistical errors added in quadrature with a 320 m s$^{-1}$ systematic error. }
\end{deluxetable}

\begin{deluxetable}{lcc}
\tabletypesize{\scriptsize}
\tablecaption{Orbital and System Parameters}
\tablewidth{0pt}
\tablehead{
\colhead{Parameter} & \colhead{Value} & \colhead{Units}}

\startdata

$P$ & 246.73$\pm0.49$ & days\\
$\gamma$ & $-0.063\pm0.078$ & km s$^{-1}$\\
$K_{1}$ & $7.02\pm0.12$ & km s$^{-1}$\\
$e$ & $0.065\pm0.016$ & \\
$\omega$ & $177\pm17$ & $^\circ$\\
$T_{0}$ & $2453537\pm11$ & HJD\\
$a_{1}\sin{i}$ & $23.75\pm0.41$ & $10^{9}$ m\\
$M_{2}\sin{i}$ & $0.2062(M_{1}+M_{2})^{(2/3)}\pm0.0034$ & $M_{\sun}$\\
V$\sin{i}$ & 16.5$\pm0.5$ & km s$^{-1}$\\
$J$ & 13.259$\pm0.024$ & \\
$H$ & 12.535$\pm0.023$ &\\
$K_{s}$ & 12.134$\pm0.026$ &\\
RA & 03:20:28.39 & hh:mm:ss (J2000)\\
DEC & $-$04:46:36.4 & dd:mm:ss  (J2000)\\

\enddata
\tablecomments{Derived and observed parameters of the 2M0320$-$04 system.}
\end{deluxetable}

\clearpage

\begin{figure}
\epsscale{1.}
\plotone{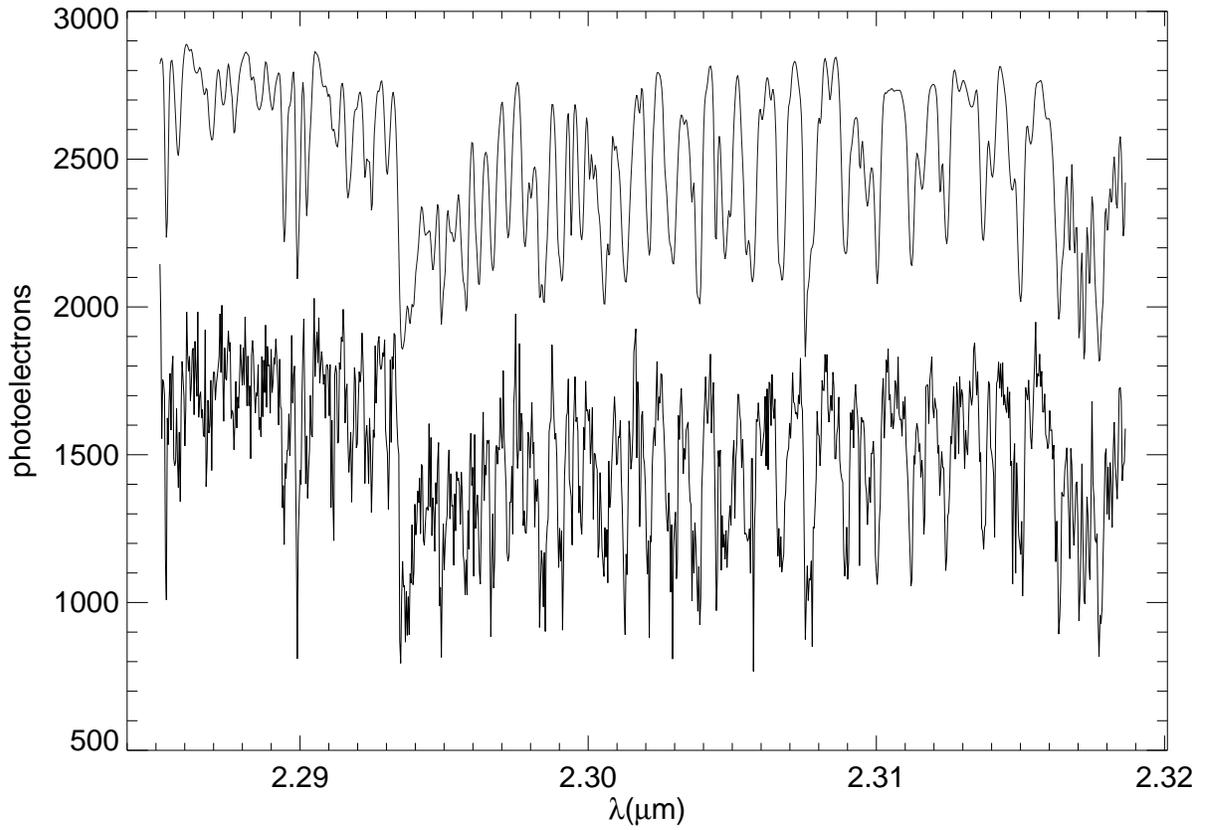}
\caption{Example of a NIRSPEC spectrum of 2M0320$-$04 (bottom) along with the best fit model offset by a constant value (top). The spectra consist of rotationally broadened CO and H$_{2}$O features from the L dwarf along with narrow telluric CH$_{4}$ lines. }\label{figc}
\end{figure}

\begin{figure}
\epsscale{1.}
\plotone{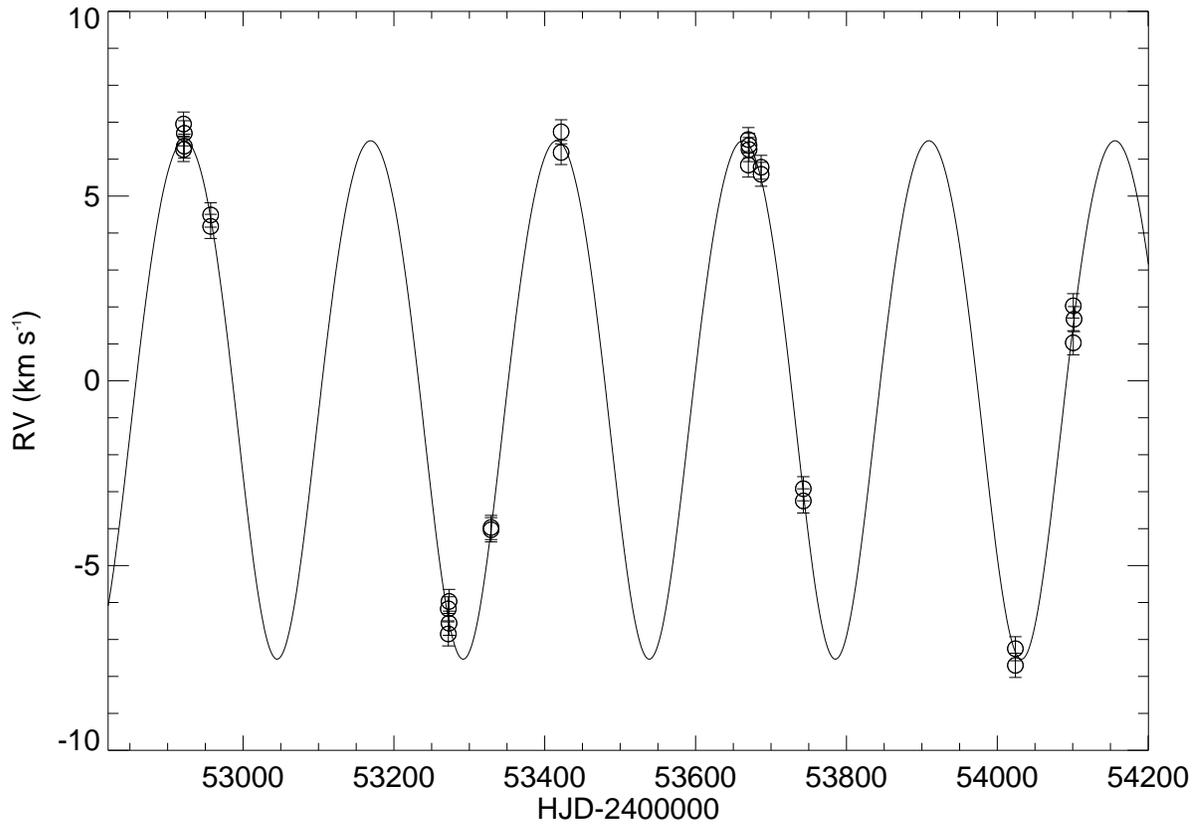}
\caption{Radial-velocity measurements and orbital solution for 2M0320$-$04 plotted as a function of time. Our observations span approximately five orbital periods.}\label{figb}
\end{figure}

\end{document}